\documentclass[conference]{IEEEtran}
\IEEEoverridecommandlockouts
\usepackage{cite}
\usepackage{amsmath,amssymb,amsfonts,amsthm}
\usepackage{dsfont}
\usepackage{graphicx}
\usepackage{xcolor}
\usepackage{booktabs}
\usepackage{multirow}
\usepackage[hidelinks]{hyperref}

\newtheorem{proposition}{Proposition}
\newtheorem{lemma}{Lemma}

\AtBeginDocument{\setlength{\abovedisplayskip}{4pt plus 2pt minus 2pt}\setlength{\belowdisplayskip}{4pt plus 2pt minus 2pt}\setlength{\abovedisplayshortskip}{2pt}\setlength{\belowdisplayshortskip}{3pt plus 1pt}}

\newcommand{\SINR}{\mathrm{SINR}}

\newcommand{\new}[1]{#1}
\newcommand{\dB}{\,\mathrm{dB}}

\begin{document}

\title{Missed Interferers in Ray-Traced Radio Maps: A Digital-Twin and Stochastic Geometry Study of Aerial Coverage}

\author{\IEEEauthorblockN{Saad Masrur and \.{I}smail G\"{u}ven\c{c}}
\IEEEauthorblockA{Department of Electrical and Computer Engineering, North Carolina State University, Raleigh, NC, USA\\
Email: \{smasrur, iguvenc\}@ncsu.edu}
\thanks{This work was supported by the NASA University Leadership Initiative (ULI) under Award 80NSSC25M7102.}}

\maketitle

\begin{abstract}
Ray-traced digital twins are increasingly used to plan where cellular-connected uncrewed aerial vehicles (UAVs) can fly. In this paper, we build a Sionna RT digital twin \new{around Raleigh, NC,} with 18,591 buildings and two overlapping sub-1~GHz operator networks\new{,} and study the downlink signal-to-interference-plus-noise ratio (SINR) of UAVs at 40--300~m. We show that the Monte Carlo radio-map solver of Sionna RT misses distant line-of-sight (LOS) interferers on aerial planes. With $10^7$ rays per sector and 5~m cells, it therefore underestimates the coverage at low SINR thresholds and overestimates it at high thresholds, by up to 26 percentage points at $\pm2$~dB and 50--100~m. Using a LOS-aware deterministic engine \new{validated with} Sionna's path solver, we then develop a stochastic geometry model with strongest-sector association and own-site leakage that matches the engine without fitting. The analysis shows that the interference grows logarithmically with the modeled network radius, so a twin that omits distant LOS sites overestimates the coverage, and that altitude mainly \new{reduces the coverage at high SINR thresholds}. Finally, selecting the better of two operators raises the minimum SINR along the best corridor by 1.0--3.9~dB up to 100~m.
\end{abstract}


\section{Introduction}
Terrestrial cellular networks are a promising way to connect uncrewed aerial vehicles (UAVs) beyond visual line of sight without dedicated infrastructure~\cite{zeng2019cellular}. Since base station (BS) antennas are tilted toward the ground and an airborne receiver sees many BSs in line-of-sight (LOS), the aerial link is limited by interference rather than by signal strength~\cite{3gpp36777}, \cite{geraci2018understanding}\new{. Therefore,} a UAV corridor is usable only where the signal-to-interference-plus-noise ratio (SINR) is high enough~\cite{hossen2026beyond}. Flight measurements at the NSF Aerial Experimentation and Research Platform for Advanced Wireless (AERPAW) show that both the reference signal received power (RSRP) and the SINR drop by about 4~dB from 30~m to 50~m~\cite{singh2026modeling}.

Ray tracers such as Sionna RT~\cite{hoydis2023sionna,aoudia2025sionna} make it practical to import a city, place the operators' sites, and compute radio maps at several altitudes, as done for UAVs at AERPAW~\cite{lee2025analysis}. However, the radio-map solver estimates the cell power by Monte Carlo ray sampling. In ray launching, the ray spacing is known to grow with distance~\cite{durgin1997advanced}, but, to our knowledge, how the number of rays and the cell size affect interference-limited metrics on aerial planes has not been quantified, nor how much of the surrounding network a twin must include.

Stochastic geometry (SG) models~\cite{andrews2011tractable,haenggi2012stochastic} show that UAV coverage drops sharply at high altitude unless the BS tilt is changed~\cite{azari2019cellular}\new{. Aerial} corridor studies optimize the BS uptilt~\cite{maeng2023uptilt} or the tilts and powers of the whole network~\cite{karimi2023analysis}\new{,} and show that strongest-signal association outperforms nearest-BS association~\cite{maeng2024corridor}. However, these studies design networks or assume idealized ones and are rarely compared with a site-specific twin of a real network, while the twins themselves are rarely checked against an analysis.

In this paper, we study what \new{an aerial corridor digital twin (ACDT)} must model to predict UAV SINR coverage, using a Sionna RT \new{ACDT} of two 600~MHz networks around NC State and AERPAW. We 1) derive the ray-hit statistics of Monte Carlo radio maps on aerial planes, which apply the ray-spacing argument of ray launching to interference-limited coverage, and quantify their bias for $10^6$--$10^8$ rays and 5--20~m cells; 2) validate a LOS-aware deterministic engine against Sionna's path solver; 3) develop an SG model with strongest-sector association and own-site leakage that needs no fitting, and derive a logarithmic law for LOS interference; and 4) quantify the effects of altitude, association, load, and \new{cellular provider (operator)} selection.

\section{Digital Twin and System Model}\label{sec:model}
\subsection{Scene and Networks}
\begin{figure}[t]
\centering
\includegraphics[width=\columnwidth]{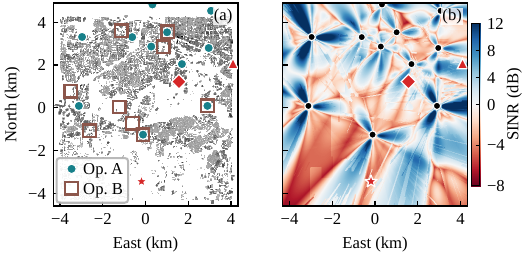}
\caption{\new{The ACDT}: (a) building heights (gray) and the sites of the two operators over the analysis area; (b) downlink SINR of Operator~A at $h=100$~m from \new{DT-det} (Section~\ref{sec:rm}; sites as dots). \new{Red markers: downtown Raleigh (triangle), NC State Centennial Campus (diamond), and the AERPAW Lake Wheeler Field Labs (star).}}
\label{fig:twin}
\end{figure}

We consider the area \new{ranging} from downtown Raleigh, NC, to the AERPAW testbed \new{at the Lake Wheeler Field Labs}, which includes the NC State campus, as shown in Fig.~\ref{fig:twin}(a). The scene was built in Blender from OpenStreetMap and exported to Sionna RT~\cite{hoydis2023sionna}. It contains 18,591 buildings whose walls and roofs use the ITU marble and metal materials, respectively, on a flat ITU concrete ground that also represents the roads. Two operator networks are placed at their publicly listed site locations: Operator~A with 11 sites at 635~MHz and Operator~B with 9 sites at 617~MHz. Every site has three sectors at azimuths $0^\circ$, $120^\circ$, and $240^\circ$, each with a $4\times4$ half-wavelength panel of 3GPP TR~38.901 elements~\cite{3gpp38901} and uniform precoding. Each sector transmits 41~dBm from a height of $h_{\mathrm{B}}=30$~m with a $6^\circ$ mechanical downtilt. The $8.64\times9.50$~km analysis area covers all buildings and sites; at ten UAV altitudes $h\in\{40,50,60,75,100,125,150,200,250,300\}$~m, it is divided into $5\times5$~m map cells (cells hereafter), and the SINR is evaluated for every cell. In Fig.~\ref{fig:twin}(b), we show the resulting SINR map of Operator~A at 100~m: high SINR is found only along the main lobes of nearby sectors, and large parts of the airspace lie between $-4$ and 0~dB. The building heights in the scene have a median of 7.3~m and a 99th percentile of 20.1~m, and only seven buildings exceed the 30~m antenna height. Consequently, ray tests on the scene show that 99--100\% of Operator~B links and 93--96\% of Operator~A links are LOS for $h\geq40$~m; the Operator~A shortfall comes almost entirely from one site next to a taller building. Hence, above the rooftops, the propagation is close to free space.

\subsection{Signal Model}
Let $\mathcal{K}$ denote the set of sectors of an operator and $P_k(\mathbf{x})$ the power received from sector $k$ at the UAV position $\mathbf{x}$. For a LOS link, we have
\begin{equation}\label{eq:pk}
P_k(\mathbf{x}) = P_{\mathrm{t}}\,G_k(\mathbf{u}_k)\left(\frac{\lambda_{\mathrm{c}}}{4\pi d_k}\right)^{2},
\end{equation}
where $P_{\mathrm{t}}$ is the sector power, \new{$G_k(\cdot)$ is the sector gain, $\mathbf{u}_k$ is the departure direction toward the UAV, $\lambda_{\mathrm{c}}$ is the wavelength, and $d_k$ is the link length}; the UAV antenna is isotropic. The \new{sector} gain is the product of the element gain $G_{\mathrm{E}}(\theta',\varphi')$ of~\cite{3gpp38901}, which has an 8~dBi peak and $65^\circ$ beamwidths, and the array factor \new{$G_{\mathrm{AF}}(\mathbf{u}')$} of the uniformly weighted panel\new{, where $(\theta',\varphi')$ and $\mathbf{u}'=(u'_x,u'_y,u'_z)$ give the direction in the tilted sector frame,} \new{and}
\begin{equation}\label{eq:af}
G_{\mathrm{AF}}(\mathbf{u}') = \frac{1}{16}\left[\frac{\sin(2\pi u'_y)}{\sin(\pi u'_y/2)}\right]^{2}\left[\frac{\sin(2\pi u'_z)}{\sin(\pi u'_z/2)}\right]^{2}.
\end{equation}
The UAV is served by the strongest sector $k^\star=\arg\max_k P_k(\mathbf{x})$, and its SINR is
\begin{equation}\label{eq:sinr}
\SINR(\mathbf{x}) = \frac{P_{k^\star}(\mathbf{x})}{\rho\sum_{k\neq k^\star}P_k(\mathbf{x})+\sigma^2},
\end{equation}
where $\rho\in(0,1]$ is the average activity (load) factor of the interfering sectors ($\rho=1$ unless stated) and $\sigma^2=-100$~dBm is the noise power for 5~MHz and a 7~dB noise figure. The coverage at threshold $T$ is the fraction of the airspace at altitude $h$ where $\SINR(\mathbf{x})\geq T$, \new{i.e.,}
\begin{equation}\label{eq:cov}
C(T,h)=\frac{1}{|\mathcal{A}|}\int_{\mathcal{A}}\mathds{1}\{\SINR(\mathbf{x})\geq T\}\,\mathrm{d}\mathbf{x},
\end{equation}
\new{where $\mathcal{A}$ is the analysis area at altitude $h$ and $|\mathcal{A}|$ its area. In the ACDT, the integral becomes the fraction of the $N_{\mathrm{c}}$ map cells whose centers $\mathbf{x}_i$ satisfy $\SINR(\mathbf{x}_i)\geq T$.} We focus on thresholds between $-4$~dB and $+4$~dB, which correspond to the low-rate (60--100~kb/s) links of UAV command and control~\cite{3gpp36777}, \new{and use $-2$~dB and $+2$~dB as representative low and high thresholds}.

\section{Ray Sampling of Aerial Radio Maps}\label{sec:rm}
In this section, we derive the ray-hit statistics of the Sionna RT radio-map solver on an aerial plane\new{, validate a deterministic engine, and quantify the radio-map errors}.

\subsection{Ray-Hit Statistics}
The radio-map solver divides the measurement plane at altitude $h$ into square cells of area $a$ and computes one path gain per cell and transmitter. Each transmitter emits $N_{\mathrm{s}}$ rays nearly uniformly over the sphere, and the path gain of a cell is estimated from the power carried by the rays that cross it~\cite{aoudia2025sionna}. This estimate is unbiased, but its variance is large. A cell that no ray crosses reports zero power, whereas a cell crossed by a single ray reports a \new{scaled version of the true received power}. Since the SINR is a nonlinear function of the received powers, its statistics are biased when $N_{\mathrm{s}}$ is too small. The following result gives the probability that a cell receives no ray.

\begin{proposition}\label{prop:hits}
For a transmitter at \new{the BS} height $h_{\mathrm{B}}$ and cells of area $a$ on a plane at height $h>h_{\mathrm{B}}$, the mean number of direct and ground-reflected rays crossing a cell at horizontal distance $r$ is
\begin{equation}\label{eq:nbar}
\bar{n}(r)=\frac{N_{\mathrm{s}}a}{4\pi}\left[\frac{\Delta h}{(r^2+\Delta h^2)^{3/2}}+\kappa\frac{h+h_{\mathrm{B}}}{\left(r^2+(h+h_{\mathrm{B}})^2\right)^{3/2}}\right],
\end{equation}
where $\Delta h=h-h_{\mathrm{B}}$ and $\kappa\in[0,1]$ is the fraction of unobstructed specular ground reflections. Under a Poisson approximation of the quasi-uniform ray lattice, the cell receives no ray with probability $e^{-\bar{n}(r)}$.
\end{proposition}
\begin{proof}
Seen from the transmitter, a cell at slant distance $d$ covers the solid angle $a\cos\new{\gamma}/d^2=a\,\Delta h/d^3$, where $\new{\gamma}$ is the angle of incidence on the plane. Since the $N_{\mathrm{s}}$ rays are spread uniformly over the full solid angle $4\pi$, the cell receives on average $N_{\mathrm{s}}a\,\Delta h/(4\pi d^3)$ direct rays, which is the first term in (\ref{eq:nbar}). Ground-reflected rays behave like direct rays from an image transmitter at height $-h_{\mathrm{B}}$, which gives the second term with $\Delta h$ replaced by $h+h_{\mathrm{B}}$.
\end{proof}

Requiring that a cell at distance $r_{\max}$ receives at least one ray with probability $1-\epsilon$, i.e., $e^{-\bar{n}(r_{\max})}\leq\epsilon$, and keeping only the direct term\new{,} gives the minimum number of rays\new{:}
\begin{equation}\label{eq:ns}
N_{\mathrm{s}}\geq \frac{4\pi\ln(1/\epsilon)\,(r_{\max}^2+\Delta h^2)^{3/2}}{a\,\Delta h}.
\end{equation}
For $r_{\max}=6$~km, $a=25$~m$^2$, $\Delta h=20$~m, and $\epsilon=0.05$, (\ref{eq:ns}) requires $1.6\times10^{10}$ rays per sector, which is 1,600 times our baseline of $10^7$ and 160 times $10^8$. Since $\bar{n}\propto a\,\Delta h$, larger cells, which blur the map, and higher planes receive more rays, so both sampling errors shrink with altitude.

\subsection{LOS-Aware Deterministic Engine \new{(DT-det)}}
Since aerial links are mostly LOS, we evaluate the received powers deterministically on the same scene, grid\new{,} and BS configuration. For every site and grid point, the engine performs three steps. First, it tests the visibility of the direct path with ray casts on the scene geometry through Mitsuba, the ray-tracing kernel of Sionna RT, and evaluates (\ref{eq:pk}) for the three sectors. Second, it adds the specular ground reflection with the Fresnel coefficient of the concrete ground. Finally, it applies a single knife-edge loss of ITU-R P.526~\cite{itu526}, computed on a 2~m surface model rasterized from the same scene, to the few obstructed paths. Hence, the engine determines the visibility of every link exactly but neglects building reflections and scattering. Path powers are added incoherently as in the radio-map definition, and three rays are cast per site and grid point, one for the direct path and two for the ground reflection. We refer to the results of this engine as DT-det.

\new{We validate DT-det against Sionna's path solver. Unlike DT-det, the path solver launches a full set of rays from each sector, traces the individual LOS and specularly reflected paths, here up to depth two and without diffraction, to a receiver at the cell center, and sums their powers incoherently.} For 100 random cells per operator and altitude (A at 50, 100, and 200~m; B at 100~m), the LOS decisions agree for all links, the median SINR error is below 0.2~dB, the root-mean-square error (RMSE) is 1.1--1.4~dB, and the 0~dB coverage of the same cells differs by at most 5\new{\%}; the residual errors come from building reflections. \new{However, the path solver needs about 3~s per point on our 2-core CPU (Intel Xeon, 2.1~GHz), whereas DT-det evaluates all 3.3 million cells of an altitude in about 2~min.}

\subsection{Radio-Map Errors}
\begin{figure}[t]
\centering
\includegraphics[width=\columnwidth]{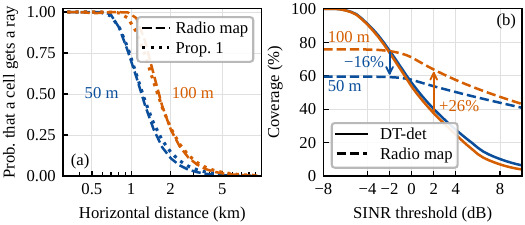}
\caption{Radio maps of Operator~A ($10^7$ rays per sector, 5~m cells) at \new{corridor heights of} 50~m and 100~m. (a) Probability that a cell receives a ray from a sector vs. horizontal distance: radio map (dashed) and Proposition~\ref{prop:hits} with empirical $\kappa=0.56$ (dotted). (b) Coverage: DT-det (solid) and radio map (dashed)\new{, with errors at $\pm2$~dB}.}
\label{fig:rm}
\end{figure}

\new{In Fig.~\ref{fig:rm}, we show radio-map results of Sionna RT with $10^7$ rays per sector.} In Fig.~\ref{fig:rm}(a), we show the probability that a 5~m cell receives at least one ray from a sector as a function of its horizontal distance from that sector. At 50~m, only 60\% of the cells 1.0--1.2~km away and 7\% of those 2.2--2.5~km away receive any ray, although nearly all of them are in LOS. Proposition~\ref{prop:hits} matches the simulation within 0.09 at both altitudes with a single empirical constant $\kappa=0.56$, which also absorbs building reflections and the \new{ray lattice}.

In Fig.~\ref{fig:rm}(b), we \new{compare the SINR coverage of Operator~A from the radio map with that from DT-det. The DT-det curves for 50~m and 100~m nearly overlap, so the gaps between the dashed and solid curves are sampling errors of the radio map.} The radio map makes two errors. First, 40\% of the cells at 50~m and 24\% at 100~m receive no ray from any sector. These cells count as outages, so the radio map underestimates the coverage at low thresholds; at $-2$~dB and \new{a corridor height of} 50~m, it gives \new{a coverage of} 59.1\%, whereas DT-det gives 74.9\%. Second, the cells that receive rays miss most distant interferers, \new{so their SINR is too high and} the radio map overestimates the coverage at high thresholds; at $+2$~dB and 100~m, it gives 63.7\%, whereas DT-det gives 37.5\%. Operator~B shows \new{similar} errors. Moreover, since fewer cells are empty at 100~m, the \new{SINR coverage} of the radio map increases from 50~m to 100~m at both thresholds, whereas that of DT-det decreases slightly. Hence, a radio map with $10^7$ rays per sector not only misestimates the coverage but also reverses its altitude trend. \new{Differences between coverage values are absolute and given in \% throughout the paper.}

\begin{table}[t]
\caption{Effect of the ray budget and cell size of the radio map at \new{a corridor height of} $h=100$~m (Operator~A)\new{. $C(T)$: coverage at threshold $T$; $\Delta I_{10}$: 10th percentile of the interference error; $C_I$: $C$ with DT-det serving power and radio-map interference.}}
\label{tab:val}
\centering
\footnotesize
\setlength{\tabcolsep}{2.2pt}
\begin{tabular}{@{}lrrrrrrr@{}}
\toprule
Map & Rays & Empty & Far & $\Delta I_{10}$ & $C(-2\dB)$ & $C(+2\dB)$ & $C_I(+2\dB)$ \\
cell & /sector & (\%) & hits (\%) & (dB) & (\%) & (\%) & (\%) \\
\midrule
5~m & $10^6$ & 64.0 & 0.2 & $<\!-40$ & 35.9 & 33.2 & 98.2 \\
5~m & $10^7$ & 24.2 & 2.1 & $-35.9$ & 74.7 & 63.7 & 89.8 \\
5~m & $10^8$ & 2.7 & 18.8 & $-20.6$ & 91.5 & 62.6 & 63.1 \\
10~m & $10^7$ & 9.1 & 8.0 & $-31.4$ & 87.9 & 66.5 & 75.5 \\
20~m & $10^7$ & 1.0 & 27.8 & $-14.9$ & 91.2 & 58.3 & 55.7 \\
\multicolumn{2}{@{}l}{DT-det$^\dagger$} & 0 & 100 & 0 & 73.7 & 37.5 & 37.5 \\
\bottomrule
\multicolumn{8}{@{}l}{$^\dagger$Averaging over 10 or 20~m cells changes DT-det by $<0.1$\%.}
\end{tabular}
\end{table}

\new{Table~\ref{tab:val} shows how the two errors depend on the ray budget and the cell size at 100~m. More rays quickly remove the empty cells, but the run time grows from 20~s at $10^6$ rays to 31~min at $10^8$. With $10^7$ rays, the coverage at $-2$~dB is close to that of DT-det only because the two errors cancel. With $10^8$ rays or 20~m cells, the empty cells nearly vanish, but the coverage at $+2$~dB is still 25\% and 21\% too high.}

\new{The reason is that distant interferers are still missed: with $10^8$ rays, 81\% of the sector--cell pairs beyond 3~km receive no ray. The median interference error is then only $-2.2$~dB, but in 10\% of the cells, more than 20~dB of interference is missing ($\Delta I_{10}$), so these cells appear to have a high SINR. Indeed, the DT-det serving power combined with the radio-map interference reproduces the overestimate ($C_I$). Two further seeds change the coverage by less than 0.01\%, so these errors are systematic.}

\new{In contrast, at 300~m with 20~m cells, where (\ref{eq:ns}) is nearly met, $10^8$ rays reach 99\% of the far sector--cell pairs, and the radio map agrees with DT-det within 3\% at $\pm2$~dB\new{; its median interference is 0.6~dB higher, consistent with the building reflections that DT-det neglects}. Hence, the errors stem from the ray budget, not from Sionna RT itself, which we use in version 1.2.1 with a maximum depth of five, specular reflection, refraction, and wedge diffraction. For the SINR, radio maps of aerial planes should be used only when (\ref{eq:ns}) is satisfied or after a convergence check.}

\section{Stochastic Geometry Model}\label{sec:sg}
In this section, we develop a stochastic geometry (SG) model of the coverage defined in (\ref{eq:cov}). \new{Unlike SOSGAD~\cite{maeng2025altitude}, which models the mean aggregate received power at a UAV with fixed antenna gains and compares it with ray tracing for the total power only, our model gives the SINR coverage with 3D three-sector patterns, strongest-sector association, and own-site leakage.} We consider a UAV at altitude $h$ and a disc of radius $R$ centered at it, and we model the sites of an operator within this disc as a homogeneous Poisson point process (PPP) $\Phi$ of density $\lambda$. As in the \new{ACDT}, each site has three sectors with gain $G_k$ and LOS links as in (\ref{eq:pk})\new{, i.e., free-space propagation with a path-loss exponent of two}; however, each site has a random orientation, and neither building blockage nor the ground reflection is modeled. A site is described by its horizontal distance $r$ and azimuth $\phi$ relative to the UAV, with $\phi$ measured from the boresight of its first sector. It contributes two powers,
\begin{equation}\label{eq:YZ}
Y(r,\phi)=\frac{P_{\mathrm{t}}\beta\,G_{\max}(r,\phi)}{r^2+\Delta h^2},\quad Z(r,\phi)=\frac{P_{\mathrm{t}}\beta\,G_{\Sigma}(r,\phi)}{r^2+\Delta h^2},
\end{equation}
where $\beta=(\lambda_{\mathrm{c}}/4\pi)^2$, $G_{\max}$ is the largest \new{3D antenna gain (element gain times array factor)} of the three sectors toward the UAV, and $G_{\Sigma}$ is \new{the sum of the three sector gains}. If the site serves the UAV, $Y$ is the useful power and $Z-Y$ is interference; otherwise, all of $Z$ is interference. The UAV is served by the site with the largest $Y$, and the following lemma gives its location.

\begin{lemma}\label{lem:serv}
The best-sector powers $\{Y_x\}_{x\in\Phi}$ form a PPP on $\mathbb{R}_+$ with intensity measure
\begin{equation}\label{eq:Lambda}
\Lambda(y)=\lambda\int_{0}^{2\pi}\!\!\int_{0}^{R}\mathds{1}\{Y(r,\phi)>y\}\,r\,\mathrm{d}r\,\mathrm{d}\phi,
\end{equation}
and the serving site is at $(r,\phi)$ with density $\lambda r\,e^{-\Lambda(Y(r,\phi))}$.
\end{lemma}
\begin{proof}
The first statement follows from the marking and mapping theorems~\cite{haenggi2012stochastic}, with the site orientations as independent marks. The second holds because the site at $(r,\phi)$ serves exactly when no other site has a larger $Y$, a void event of probability $e^{-\Lambda(Y)}$.
\end{proof}

Given that the serving site has best-sector power $s$, every other site has $Y<s$, so the interfering sites form a PPP restricted to the region $\{Y<s\}$. By Campbell's theorem, the mean and variance of their total power are
\begin{equation}\label{eq:mv}
\mu(s)=\lambda\!\iint_{Y<s}\!Z\,r\,\mathrm{d}r\,\mathrm{d}\phi,\quad v(s)=\lambda\!\iint_{Y<s}\!Z^2\,r\,\mathrm{d}r\,\mathrm{d}\phi .
\end{equation}
Approximating this interference by a Gamma variable with the same mean and variance, as in~\cite{heath2013modeling}, yields the coverage.

\begin{proposition}\label{thm:cov}
The coverage probability of a UAV at altitude $h$ is approximately
\begin{equation}\label{eq:thm}
C(T,h)\approx\lambda\int_{0}^{2\pi}\!\!\int_{0}^{R}e^{-\Lambda(Y)}\,F_{\Gamma}\!\left(\Psi;\frac{\mu^2}{v},\frac{v}{\mu}\right)r\,\mathrm{d}r\,\mathrm{d}\phi,
\end{equation}
where $\Psi=[Y(T^{-1}-\rho\,\xi)-\sigma^2]/\rho$, $Y=Y(r,\phi)$, $\mu=\mu(Y)$, $v=v(Y)$, $\xi(r,\phi)=G_{\Sigma}/G_{\max}-1$ is the \new{sum of the linear gains of the other two sectors of the serving site, normalized by the serving gain}, and $F_{\Gamma}(x;\nu,\vartheta)$ is the Gamma distribution function with shape $\nu$ and scale $\vartheta$, with $F_{\Gamma}=0$ for $x\leq0$.
\end{proposition}
\begin{proof}
Condition on the serving site at $(r,\phi)$\new{, whose density is given by Lemma~\ref{lem:serv}. Its best sector delivers $Y$, and its other two sectors deliver the interference $\xi Y$. With $I$ the interference from all other sites, (\ref{eq:sinr}) gives $\SINR=Y/(\rho(\xi Y+I)+\sigma^2)$, so $\SINR\geq T$ if and only if $I\leq\Psi$. With the mean and variance in (\ref{eq:mv}), $\Pr\{I\leq\Psi\}\approx F_{\Gamma}(\Psi;\mu^2/v,v/\mu)$, and averaging over $(r,\phi)$ gives (\ref{eq:thm}).}
\end{proof}

Proposition~\ref{thm:cov} \new{takes a few seconds to evaluate}. Only the site density and the disc radius are taken from the \new{ACDT}, and no parameter is fitted. For a stationary PPP, this coverage probability equals the covered area fraction~\cite{andrews2011tractable}, which is the quantity computed in DT-det; since the \new{ACDT} has edges, the equality is approximate.

\new{The SG model also reveals an SINR ceiling. Even without other sites, the other two sectors of the serving site deliver $\xi$ times the serving power, so this site cannot provide an SINR above $1/(\rho\xi)$. With the full 3D patterns, not a fixed sidelobe level, $\xi$ depends on the UAV position: $\xi\ll1$ in the main lobe of the serving sector, but $\xi\approx2$ directly above a site, where all three sectors have similar sidelobe gains, and $\xi\approx1$ between two sectors, which gives ceilings of $-3$~dB and 0~dB at full load. There, the strongest sector often belongs to a farther site whose main lobe points toward the UAV (Section~\ref{sec:results}-C).}

As known for a path-loss exponent of two~\cite{haenggi2012stochastic}, the interference of an infinite LOS network diverges. Proposition~\ref{prop:log} quantifies it for three-sector sites with 3D panel patterns, which determines how large a twin must be.

\begin{proposition}\label{prop:log}
Consider the sites in the ring $R_1<r<R_2$ around the UAV. The mean total interference from all their sectors is
\begin{equation}\label{eq:ann}
\bar{I}(R_1,R_2)=2\pi\lambda P_{\mathrm{t}}\beta\int_{R_1}^{R_2}\frac{\bar{G}_{\Sigma}(r)\,r}{r^2+\Delta h^2}\,\mathrm{d}r,
\end{equation}
where $\bar{G}_{\Sigma}(r)$ is the \new{sum of the 3D antenna gains of the three sectors} of a site toward the UAV, averaged over the site orientation, at the elevation angle $\arctan(\Delta h/r)$. Far from the UAV ($R_1\gg\Delta h$), the UAV is seen near the horizon, so $\bar{G}_{\Sigma}(r)$ approaches $\bar{G}_{\mathrm{hor}}$, the mean total gain of a site toward the horizon, and
\begin{equation}\label{eq:log}
\bar{I}(R_1,R_2)\approx\pi\lambda P_{\mathrm{t}}\beta\,\bar{G}_{\mathrm{hor}}\ln\frac{R_2^2+\Delta h^2}{R_1^2+\Delta h^2}.
\end{equation}
\end{proposition}
\begin{proof}
Equation (\ref{eq:ann}) is Campbell's theorem applied to $Z$ in (\ref{eq:YZ}), and (\ref{eq:log}) follows by replacing $\bar{G}_{\Sigma}(r)$ with $\bar{G}_{\mathrm{hor}}$.
\end{proof}

Proposition~\ref{prop:log} has two consequences. First, \new{for a fixed inner radius $R_1$, every doubling of the outer radius $R_2$ adds the same mean interference, $2\pi\lambda P_{\mathrm{t}}\beta\bar{G}_{\mathrm{hor}}\ln2$. In plain words, the sites between 4 and 8~km add as much interference as those between 2 and 4~km, because the number of sites grows with $r^2$, whereas the power of each falls with $r^{-2}$.} Hence, far sites cannot be ignored, unlike for path-loss exponents above two. In practice, buildings, terrain, and foliage block part of the distant links, so (\ref{eq:log}) is an upper bound on the far interference. For a three-sector site with our panel, \new{averaging $G_{\Sigma}$ over the site orientation at zero elevation gives} $\bar{G}_{\mathrm{hor}}=12.7$~dBi, and (\ref{eq:log}) is within 2\% of (\ref{eq:ann}) for $R_1\geq4$~km at $h\leq100$~m. Second, since $h$ enters (\ref{eq:log}) only through $\Delta h\ll R_1$, the far interference hardly depends on altitude, whereas the serving power changes as $\partial\ln Y/\partial h=-2\Delta h/(r_{\mathrm{s}}^2+\Delta h^2)+\partial\ln G_{\max}/\partial h$, where $r_{\mathrm{s}}$ is the serving distance. Hence, climbing mainly weakens the serving signal close to a site, where the SINR is high, whereas locations far from all sites hardly change.

\section{Numerical Results}\label{sec:results}
In this section, we use DT-det, not the radio map, \new{because the radio map is biased at feasible ray budgets (Section~\ref{sec:rm}),} and the SG model to study the effect of altitude, network radius, association, load, and operator diversity. Unless stated otherwise, solid lines show DT-det, dotted lines analytical results, dashed lines other simulations, and colors the altitude\new{. The} SG model sets $\lambda$ to the number of sites of each operator divided by the analysis area, and $R=5.11$~km, so that the disc has the same area as the \new{ACDT}. All DT-det results include only the sites in the analysis area and are thus optimistic if the real network extends beyond it (Section~\ref{sec:results}-B). Hence, the absolute coverage values are a reference case, whereas the radio-map errors, the own-site leakage, and the logarithmic law do not depend on the exact layout.

\subsection{Effect of Altitude}
\begin{figure}[t]
\centering
\includegraphics[width=\columnwidth]{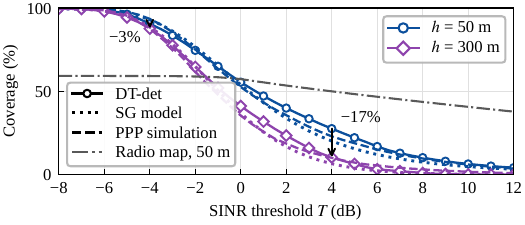}
\caption{Coverage of Operator~A versus SINR threshold at 50 and 300~m: DT-det (solid), SG model (dotted), PPP simulation without ray tracing (dashed)\new{, and radio map with $10^7$ rays at 50~m (gray dash-dotted)}. The arrows mark the 50-to-300~m change at $\pm4$~dB.}
\label{fig:covT}
\end{figure}

In Fig.~\ref{fig:covT}, \new{the SG model matches DT-det without any fitting. Over all ten altitudes and thresholds from $-6$~dB to 10~dB in 0.5~dB steps, the RMSE is 3.9\% (3.2\% for Operator~B). Against a Monte Carlo simulation of the same PPP network, it is only 2.0\%, so most of the error comes from the real site layout. With random site azimuths, the DT-det coverage at 0~dB and 100~m changes by \new{at most 3.6\% for Operator~A and 4.7\% for Operator~B}. The radio map (gray) deviates from both by up to 32\%.} Even at 50~m, only 56\% of the airspace reaches 0~dB, because nearly all 33 sectors are in LOS and many arrive with comparable powers. \new{Altitude mainly affects the high thresholds. From 50~m to 300~m, the coverage falls by only 3\% at $-4$~dB but by 17\% at $+4$~dB (arrows). As predicted by Proposition~\ref{prop:log}, the median serving power falls by 3.1~dB as the UAV rises out of the main lobe of a nearby sector, whereas the median interference falls by 0.9~dB. Hence, low-threshold ($\leq0$~dB) command-and-control links can fly at any altitude within 40--150~m, where their coverage changes by at most 3.3\%, whereas high-SINR links should fly low.}

\subsection{Effect of the Modeled Network Radius}
\begin{figure}[t]
\centering
\includegraphics[width=\columnwidth]{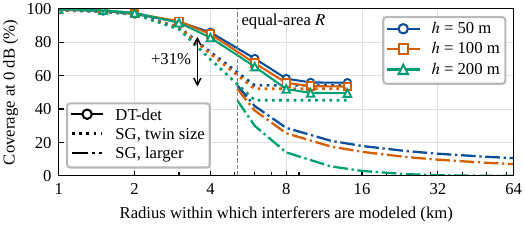}
\caption{Coverage of Operator~A at 0~dB versus the radius within which interferers are modeled: DT-det (solid), SG model for the \new{ACDT-size} network (dotted), and SG model for a larger network of radius $R$ (dash-dotted).}
\label{fig:range}
\end{figure}

In Fig.~\ref{fig:range}, we show the coverage at 0~dB when only the interferers within a given radius of the UAV are counted. \new{At 100~m, ignoring the interferers beyond 4~km overestimates the coverage by 31\% (arrow), and the error falls to about 2\% only at 8~km, \new{about the ACDT size}. The SG curve for the ACDT-size network drops faster than DT-det, because a UAV near the edge of the ACDT still sees sites up to 13~km away. If the network continues beyond the ACDT at the same density, the coverage keeps falling, from 52\% to 14\% at 16~km, as predicted by Proposition~\ref{prop:log} if these links remain LOS. Hence, the modeled radius of a twin should be reported, and the surrounding sites should be included or bounded with (\ref{eq:log}).}

\subsection{Association and Load}
\begin{table}[t]
\caption{\new{Association and load at a corridor height of 100~m.}}
\label{tab:assoc}
\centering
\footnotesize
\setlength{\tabcolsep}{4pt}
\begin{tabular}{@{}lcc@{}}
\toprule
\new{Metric (\%)} & \new{Operator~A} & \new{Operator~B} \\
\midrule
\new{Served by a site other than the nearest} & \new{62.7} & \new{67.1} \\
\new{$C(-2\dB)$, strongest sector / nearest site} & \new{73.7 / 32.6} & \new{80.8 / 30.3} \\
\new{$C(0\dB)$, $\rho=1$, DT-det / SG} & \new{53.5 / 52.1} & \new{60.1 / 61.0} \\
\new{$C(0\dB)$, $\rho=0.5$, DT-det / SG} & \new{83.1 / 85.9} & \new{89.9 / 90.3} \\
\bottomrule
\end{tabular}
\end{table}

We compare two association rules: the UAV connects either to the strongest sector, as in a real network, or to the nearest site, as often assumed in corridor studies~\cite{maeng2023uptilt}. \new{Table~\ref{tab:assoc} summarizes the results at 100~m. The UAV is served by a site other than the nearest one at about two thirds of the locations,} because the about $25^\circ$ wide azimuth beams of the $4\times4$ panel leave gaps between the sectors of a site, and the other sectors of the nearest site leak interference, as captured by $\xi$. Unlike in~\cite{maeng2024corridor}, this mismatch is quantified for a real site layout. As a result, nearest-site association \new{predicts a coverage at $-2$~dB of only 33\% instead of 74\% for Operator~A and 30\% instead of 81\% for Operator~B. The coverage also depends strongly on the load. If the interfering sectors are active only half of the time ($\rho=0.5$), the coverage at 0~dB rises by about 30\% in both DT-det and the SG model,} which suggests a gain from inter-cell coordination. \new{Since $\rho$ scales only the mean interference, random on/off activity would add variance, and always-on NR synchronization signals correspond to $\rho=1$}.

\subsection{Dual-Operator Selection Along Corridors}
\begin{figure}[t]
\centering
\includegraphics[width=\columnwidth]{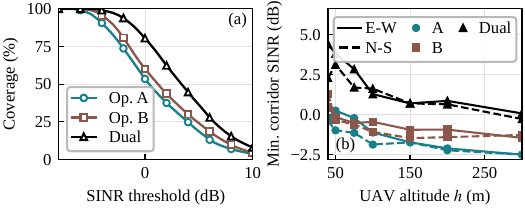}
\caption{Dual-operator selection: (a) coverage at 100~m for Operator~A, Operator~B, and the better of the two (Dual); (b) minimum SINR along the best east--west (solid) or north--south (dashed) corridor crossing the area.}
\label{fig:dual}
\end{figure}

In Fig.~\ref{fig:dual}(a), we show the coverage at 100~m when the UAV can select the better of the two operators. Multi-operator diversity is known to improve UAV reliability~\cite{colpaert2022drone}; here, dual-operator selection raises the coverage at $-2$~dB from 73.7\% for Operator~A and 80.8\% for Operator~B to 93.8\%.

In Fig.~\ref{fig:dual}(b), we show, for each altitude, the minimum SINR along the best corridor that crosses the area east--west (solid) or north--south (dashed). To obtain it, we search for the largest threshold $T^\star$ for which the cells with $\SINR\geq T^\star$ still form an eight-connected path across the area. With a single operator, $T^\star$ lies between $-2.5$~dB and 1.3~dB, depending on the operator, altitude, and direction. With dual-operator selection, $T^\star$ rises to 1.3--4.4~dB up to 100~m, which is 1.0--3.9~dB more than the better single operator. This gain is an upper bound, because the UAV changes operator at no cost. At 100~m, the best east--west and north--south corridors are 9.4~km and 11.6~km long and change operator 18 and 31 times.

\section{Conclusion}\label{sec:conc}
\new{In a Sionna RT ACDT, Monte Carlo radio maps miss distant LOS interferers on aerial planes, because the ray-hit density decays with the cube of the distance. \new{With $10^7$ rays and 5~m cells, their coverage at $\pm2$~dB is off by up to 26\%, and} more rays or larger cells remove the empty cells much faster than the missed interferers. DT-det agrees with Sionna's path solver, and an SG model reproduces DT-det without fitting. LOS interference grows logarithmically with the modeled radius \new{while distant links remain LOS}, so a twin should report its radius. Altitude mainly reduces the coverage at high thresholds, whereas association and load change it by 30--50\%, and dual-operator selection raises the minimum corridor SINR by up to 3.9~dB. As the ACDT assumes flat terrain, no foliage, \new{full Fresnel clearance,} full buffers, and assumed azimuths and tilts, its absolute numbers require field validation, such as at the NSF AERPAW platform, which is left for future work.}

\section*{Acknowledgment}
The authors used Anthropic's Claude to assist with code development, data analysis, and editing. All content was verified by the authors, who take full responsibility.

\bibliographystyle{IEEEtran}
\bibliography{refs}

\end{document}